\documentclass{aa}
\usepackage[varg]{txfonts}
\usepackage{multirow}
\usepackage{natbib}
\usepackage{hyperref}
\usepackage{booktabs} 
\usepackage{xcolor}
\usepackage{color}
\bibpunct{(}{)}{;}{a}{}{,}

\newcommand{\w}{\boldsymbol}
\newcommand{\de}{\delta}
\newcommand{\dd}{\partial} 
\newcommand{\ls}{\langle} 
\newcommand{\rs}{\rangle} 
\newcommand{\bv}{\mathbf{v}}
\newcommand{\br}{\mathbf{r}}
\newcommand{\brho}{\w{\rho}}

\begin{document}

\title{Small-scale turbulent dynamo for low-Prandtl number fluid:  Comparison of the theory with results of numerical simulations}
\titlerunning{SSD for low-$Pm$ dynamo}

\author{A.V. Kopyev\inst{1}
  \and A.S. Il'yn\inst{1,2}
     \and V.A. Sirota\inst{1}
     \and K.P. Zybin\inst{1,2}}

\offprints{A. Kopyev, \email{kopyev@lpi.ru}}

\institute{P.N.Lebedev Physical Institute of RAS, 119991, Leninskij pr.53, Moscow, Russia
  \and National Research University Higher School of Economics, 101000, Myasnitskaya 20, Moscow, Russia}

\abstract {During the past decades, significant progress has
been made in numerical simulations of the turbulent dynamo and the theoretical understanding of turbulence. However, a
quantitative
comparison between simulations and the theory of the dynamo is still lacking.} {We investigate the generation of a magnetic field by the incompressible turbulent conductive fluid near the critical regime and compare the theoretical predictions of the Kazantsev model with results of recent direct numerical simulations.
 }{The Kazantsev equation was analyzed both analytically and numerically. }{We studied  the critical magnetic Reynolds number ($Rm_c$)  and  the growth rate near the threshold 
in the limit of very high 
and for  moderate Reynolds numbers.
We argue that in the Kazantsev equation for magnetic field generation, the quasi-Lagrangian
correlator of velocities should be used
instead of the Eulerian, as is usually  implied   
when theory and simulations are compared.   The theoretical results obtained with this correlator agree well with the numerical results.
We also propose that the decrease of ~$Rm_c$  can be explained as a function of the Reynolds number  ($Re$)   
at  intermediate to high 
$Re$.   It is probably due to the Reynolds-dependent intermittency of the velocity structure function.  We show that 
the scaling exponent of this function in the inertial range strongly affects the magnetic field generation, and it is known  to be an increasing function of  the Reynolds number.}{The quasi-Lagrangian correlator in the Kazantsev theory provides results that agree well with numerical simulations. 
An ideal way to compare them is 
to find the correlator that can be substituted in the Kazantsev equation and the generation properties in the same simulation. Universal parameters at least have to be used, regardless of the properties of the pumping~scale.
The Reynolds-dependent intermittency can explain the recently observed decrease in the critical magnetic Reynolds number at small Prandtl numbers.}

\maketitle
\nolinenumbers

\section{Introduction}
The theory of how magnetic fields appear in turbulent flows of conductive fluid has wide applications, particularly in explaining the magnetic fields observed in many astrophysical objects \citep[see, e.g.,][]{brandenburg2005astrophysical, jouve2008solar, moss2013reversals, brandenburg2024resistively}.
There are two possible mechanisms to generate a small-scale magnetic field, that is, a field with a characteristic scale much smaller than 
 the integral scale of turbulence~\citep{karak2016small}.
First, there is a cascade mechanism of generation that is initiated at large scales under the influence of global shear and convective flows. Second, there is a faster generation process caused by small-scale turbulence.
The latter
has been studied in numerous papers \citep[see, e.g.,][]{zeldovich1990almighty, falkovich2001particles, brandenburg2012current} and is the subject of this paper. 

Conductive fluids can be classified by their magnetic Prandtl number, 
\begin{equation} \label{Pm}
Pm = \nu / \eta,
\end{equation}
where $\nu$ is viscosity, and $\eta$ is magnetic diffusivity. For $Pm \gg 1$, the resistive scale lies deep inside the viscous range of turbulence~\citep{batchelor1950spontaneous, zel1984kinematic, chertkov1999small}. This limit is realized, for instance,  in the interstellar medium and in processes of star formation~\citep{han2017observing}.
We concentrate on the other case of fluids with low and intermediate $Pm$. This means that the resistive scale exceeds the viscous scale, so that the excitation of magnetic fluctuations is driven by the inertial-range or  bottleneck velocity fluctuations. Low-$Pm$ processes take place in the interior of solar-type stars and planetary convective envelopes~\citep{petrovay1993origin}.

The theoretical description of the process 
in the incompressible fluid 
is based on the Kazantsev equation, which relates the evolution of the  magnetic field pair correlator and the velocity structure function. Experimental possibilities are rather limited in terrestrial conditions. Direct numerical simulations (DNS) of low $Pm$, unlike the case of high Prandtl numbers,  face  difficulties in modeling magnetic field advection in  the inertial range of turbulence
 and demonstrate some discrepancies between different simulations \citep[see, e.g.,][Fig.2]{warnecke2023numerical}.

However, during the past decades, significant progress has been made  in the theoretical understanding   of the properties of turbulence~\citep{l1997temporal, donzis2010bottleneck, biferale2011multi, iyer2020scaling}  and of the environment and technics of  DNS  \citep{iskakov2007numerical, warnecke2023numerical, rempel2023small, brandenburg2023dissipative, yeung2025small}, which has
resulted in high resolution and advanced to lower values of  $Pm$.  
This progress gives us hope that we can proceed from a qualitative correspondence to a more or less thorough quantitative comparison at least between theory and DNS. 

In the study of turbulence, after the theoretical model formulation (for decaying turbulence) in the classical  work \citep{kolmogorov1941dissipation} and its development for stationary turbulence \citep{novikov1965functionals}, the precision testing of the model results was performed in experiments \citep{moisy1999kolmogorov}. 
 The coincidence of theoretical predictions and experimental measurements  has, citing \citet{moisy1999kolmogorov}, demonstrated  'the relevance of isotropic homogeneous turbulence state approximation, used in almost all theoretical approaches to turbulence'. 
For a low-$Pm$ dynamo, such a program seems hardly possible because of experimental difficulties.  However, in the absence of experimental approbation, an accurate quantitative comparison with DNS becomes even more important. 
It could validate both of the approaches: it would verify the resolution of DNS, and verify the applicability  
of theoretical assumptions of $\delta$-time correlation \citep{Tobias_Cattaneo_Boldyrev_2012} and Gaussianity of the turbulent velocity field. It could also test nontrivial effects resulting from  non-Gaussianity, for instance, from the weakening of the magnetic field generation, and corrections for the magnetic energy spectrum \citep{kopyev2022magnetic, kopyev2022non, kopyev2024suppression}.  
Finally, if the theory were confirmed  for moderate to low magnetic Prandtl numbers, it might be scaled to extremely low $Pm$, which are unattainable by current facilities and typical for astrophysical objects.

\vspace{0.2cm}

The correct interpretation of numerical data for velocity correlators for their comparison with the  theory is one of the problems in this path.   
In the Kazantsev theory, velocity correlators are assumed to be $\delta$-correlated in time, and the velocity distribution 
acts in the Kazantsev equation   
by means of  the  multiplier $b(\rho)$ in the structure function, 
\begin{equation} \label{bKaz}  
\langle \delta \bv_{\parallel} (\brho,  t)   \delta \bv_{\parallel} (\brho,t' )  \rangle _{\rm Kaz} = 2 b(\rho) \delta (t-t')   \ , 
\end{equation}  
where the longitudinal velocity increments in two near points are 
\begin{equation}  \label{delta-v}
  \delta \mathbf{v}_{\parallel} (\boldsymbol{\rho} ,t)  =  \left(    \mathbf{v}  (\mathbf{r} +\boldsymbol{\rho},  t)  - \mathbf{v}   (\mathbf{r},  t)  \right) \frac{\boldsymbol{\rho}}{\rho}
\end{equation}
The correlator is 
independent of  $t$,  $\br$, and of the direction of $\brho$ because turbulence is homogeneous and isotropic. 

To restore the amplitude $b(\rho)$ from a given correlator, we can use the expression~\citep{vainshtein1980dynamo, kichatinov1985dynamo}
\begin{equation}  \label{b-def}  
b(\rho) = \frac 12 \int \limits_{-\infty}^{\infty} d\tau  \langle \delta \bv_{\parallel} (\brho,  \tau)   \delta \bv_{\parallel} (\brho,0 )  \rangle  .
\end{equation}
For the $\delta$ -correlated process, there is no difference  whether  $\delta \bv_{\parallel}$ 
is calculated at a fixed point $\br$ of space or at an arbitrarily moving point $\br (t)$; the correlator is the same.  
The same result is also obtained for the quasi-Lagrangian reference frame, that is,   $\mathbf{r}(t)$  tracing one arbitrary fluid particle~\citep{l1997temporal}.  This frame is distinguished, for example, by the fact that the Kazantsev predictions must be consistent with the results obtained from the evolutionary approach \citep{zel1984kinematic, chertkov1999small}  wherever both of the models are applicable; and the evolutionary approach explicitly considers the quasi-Lagrangian frame. 

In real turbulent flows, including those in DNS,  all structure functions have a nonzero correlation time, and the values of $b(\rho)$  calculated in different frames (i.e., for different $\mathbf{r}(t)$ in Eq. \ref{delta-v}) differ essentially. 
The
analysis of data from John Hopkins database shows that at high Reynolds numbers,  for all scales  $\langle \delta \bv_{\parallel} (\brho,  \tau)   \delta \bv_{\parallel} (\brho,0 )  \rangle ^{\rm Eul}  \propto 1/\tau$~(Kopyev et al. in preparation).  If this is so, the logarithmic divergence of Eq. (\ref{b-def}) corresponds to a growth of the magnetic field that is faster than exponential; this would contradict the Oseledets theorem~\citep{oseledec1968multiplicative}.
  We note that, unlike in the Eulerian case,  the quasi-Lagrangian $b(\rho)$ converges well:  
$\langle \delta \bv_{\parallel} (\brho,  \tau)   \delta \bv_{\parallel} (\brho,0 )  \rangle^{\rm Lagr}$ decays exponentially with respect to $\tau$  \citep{biferale2011multi} .  

Consider now some experiment or simulation. We determine below which $b(\rho)$ should be substituted in the Kazantsev equation in order to compare the theory results with the experiment.
 In other words, we determine which $b(\rho)$ simulates the  correct amplitude for the effective $\delta$-correlator. 
 The choice of  the  Eulerian frame ($\br = const$) in Eq. (\ref{delta-v}), although it might seem most natural, is in fact  not    preferable and is probably just incorrect. This might have been the reason that prevented \citet{mason2011magnetic}  from obtaining  a good correspondence between theory and DNS.
A consistent solution of the problem must be based on an accurate dynamical analysis and is far from being  performed yet. We assume, however, that the correct answer is to choose the quasi-Lagrangian frame. The grounds for this assumption are as follows:  \\
- First, from the renewing model \citep{zel1984kinematic}, it follows \citep{zeldovich1990almighty, rogachevskii1997intermittency} that if the correlation time is shorter than all other characteristic timescales, the quasi-Lagrangian frame gives  the correct result:  the Kazantsev approach in this case can be verified by evolutionary models.
This is even the case near the generation threshold: the characteristic time for the magnetic field evolution is long, so the $\delta$ approximation for velocity correlations works well. \\
- Second, for an arbitrary correlation time but only for high magnetic Prandtl numbers (which restricts the consideration to the viscous range of scales), the choice of the quasi-Lagrangian frame is also correct~\citep{vainshtein1980dynamo, kichatinov1985dynamo}. Recently, \citet{il2022long} have shown the quasi-Lagrangian correlators to be equivalent to effective $\delta$-correlators in long-time approximation. 

By analogy with the limit cases, it is reasonable to imply the quasi-Lagrangian velocity increments in Eqs. (\ref{delta-v}) and (\ref{b-def}) .  The relevance of this assumption is validated in this paper by means  of a direct comparison of the predictions of the theory based on the Kaznatsev equation with  the velocity structure function  $b(\rho)$  determined in this way with the results of DNS. The good agreement is the proof of our choice of the quasi-Lagrangian structure function.

\vspace{0.2cm}

We calculate the critical magnetic Reynolds number and the increments of the magnetic field  correlator near the critical regime of generation. For this purpose, we solve the Kazantsev equation with quasi-Lagrangian velocity structure function (\ref{b-def}).  The basic notations and equations are introduced in Sect.~\ref{S:2-bas-eq}.  

Unfortunately, there are not many data  on quasi-Lagrangian velocity structure functions. We therefore considered two different cases:  \\
the Taylor Reynolds number $Re_{\lambda} = 140$   (Sect.~\ref{S:3-140}), for which DNS data on the magnetic field~\citep{schekochihin2007fluctuation}  and velocity correlators~\citep{biferale2011multi, donzis2010bottleneck} are available,   \\
and  the limit of the infinite Reynolds number  (Sect.~\ref{S:4-largeRe}),   where the shape of the  velocity correlator is chosen  based on theoretical reasons. For this case, we use different  approximations and consider  the effects produced by 
 the declination  of the velocity structure function    exponent from the Kolmogorov scaling 
in the inertial range of turbulence~\citep{l1997temporal, iyer2020scaling}  and the properties of the transition range between  the  inertial and largest-eddy scales. 
We compare the results with  the numerical  data \citet{iskakov2007numerical,schekochihin2007fluctuation,brandenburg2018varying,warnecke2023numerical}  and find a quite good correspondence.  
We also show that the Reynolds-dependent intermittency of velocity structure functions \citep{iyer2020scaling} can explain the observed decrease in the critical Reynolds number at small Prandtl numbers \citep{warnecke2023numerical}. We calculate the magnetic field correlator growth rate in the vicinity of the  generation threshold and compare the results with the DNS (Sect.~\ref{S:5-incr}).

Finally, we summarize the results of the paper and  comment on  further prospects  and possible improvements in comparison of the Kazantsev theory predictions with experiments and simulations.   In particular, we indicate the properties of quasi-Lagrangian turbulence that can be found from DNS together with the magnetic generation properties, to make the comparison with the theory  quantitative (not only qualitative) and precise.

\section{Basic equations and parameters     }~\label{S:2-bas-eq}
The evolution of the magnetic field \(\mathbf{B}(\mathbf{r},t)\) is governed by the induction equation,

\begin{equation} \label{induc}
\frac{\partial \mathbf{B}(\mathbf{r},t)}{\partial t} = 
\nabla \times \bigl[ \mathbf{v}(\mathbf{r},t) \times \mathbf{B}(\mathbf{r},t) \bigr] 
+ \eta \nabla^2 \mathbf{B}(\mathbf{r},t).
\end{equation}

Since the back-reaction of the magnetic field on the flow is quadratic in field strength and the initial (seed) magnetic field is weak, we can neglect its effect on the velocity dynamics. In this kinematic regime, the magnetic field acts as a passive vector field advected by the flow.
The solenoidal velocity field \(\mathbf{v}(\mathbf{r},t)\) is treated as a prescribed stochastic field with stationary statistics.

 The correlation function of the magnetic field 
   \begin{equation}
      \label{G-def}
   G(\rho, t)=\langle  B_{\parallel}(\mathbf{r}+{\brho}, t) 
  B_{\parallel}(\mathbf{r}, t) \rangle \ ,  \ \  B_{\parallel} = \mathbf{B} \cdot \brho / \rho
  \end{equation}    
  is assumed to be independent of $\br$ in a homogenous isotropic flow. Its evolution under the assumption~(\ref{bKaz}) is described by the equation \citep{kazantsev1968enhancement}
\begin{equation}\label{E:GEq}
\frac\dd{\dd t}G(\rho, t)=2 S(\rho)  \left( G''_{\rho\rho}+
   \frac{4G'_\rho}{\rho}\right)+    2 S'
   G'_{\rho}+ 2 \left(S''+4\frac {S'}{\rho}\right)G,
\end{equation}
where
  \begin{equation*}
   S(\rho)= \eta + \frac 12  b(\rho) \ ,  \quad 
  \end{equation*}
  and the function $b(\rho) $ is  defined in (\ref{b-def}). 
We are interested in the exponential behavior of the magnetic field, and we therefore searched for a solution in the form 
\begin{equation*}
G=e^{\gamma t}  \psi(\rho) / (\rho^2  \sqrt{S} ).
\end{equation*}
Then, $\psi$ satisfies  the Schrodinger-type equation  \citep{kazantsev1968enhancement}
\begin{align}\label{eq-psi}
 & \psi''_{\rho \rho}  =  \frac {\gamma}{2S(\rho)}  \psi  +  U(\rho) \psi  \ ,  \\ \label{U-pot}
 &U =  - \frac 1{\rho^2}  \left(\frac{3\sigma(\sigma+4)+1}{4}+\frac{\rho \sigma'}{2}\right)   \ ,    \\\label{sigma}
 &\sigma(\rho)= \frac{\mathrm{d}\ln S}{\mathrm{d} \ln \rho}-1     
\end{align}
with  the boundary condition $\psi(0)=0 \ , \ \psi (\infty) <\infty$.  
We solved this equation numerically by means of a modification of the stochastic quantization idea \citep[Appendix A]{il2021evolution}.  

To compare the theory with the results of DNS, we normalized the dimensional parameters. Generally, an isotropic hydrodynamical flow is completely characterized by three dimensional parameters \citep{novikov1965functionals, moisy1999kolmogorov}: 
\begin{equation*}
\nu, \qquad   \varepsilon=\nu\left\langle \frac{\partial  v_i}{\dd x_k}\frac{\dd v_i}{\dd x_k}\right\rangle,\,\,\,\,\,\,\,u'=\sqrt{{\ls v_i v_i\rs}/{3}} = v_{\rm rms} /\sqrt{3},
\end{equation*}
where $\nu$  is the viscosity,  $\varepsilon$ is the total energy flux from larger to smaller scales, and $v_{\rm rms}$ is the volume integrated root-mean-squared velocity.
From these three parameters, we composed one universal dimensionless combination, for example,
\begin{equation}  \label{Re-lambda}
Re_\lambda\equiv\frac{u'\lambda}{\nu}=\frac{\sqrt{15}\,{u'}^{\,2}}{\sqrt{\varepsilon \,\nu}}, 
\end{equation}
where $\lambda=u'\sqrt{15\nu/\varepsilon}$ is the Taylor scale; this parameter is called the Taylor Reynolds number.  However,  the classical Reynolds number is often used,
\begin{equation*}
Re =  v_{\rm rms} L / \nu,
\end{equation*}
where $L$ is the pumping scale, or the largest-eddies scale,  and is not defined universally. This uncertainty produces difficulties in matching different experimental and/or DNS results and their comparison with the theory (e.g.,~\citet{brandenburg2018varying} reported   some obstacles 
when comparing their results with those of \citealt{schekochihin2007fluctuation}). We defined 
\begin{equation}  \label{Re-Schek}
Re^{\rm Sch} =\frac{Re_\lambda^2}{30}  =\frac{{u'}^{\,4}}{2 \varepsilon \,\nu}.
\end{equation}
This corresponds to the relations between $Re $ and $Re_{\lambda}$  used by \citet{schekochihin2007fluctuation}.

The magnetic properties of the fluid can be described by the dimensionless magnetic Prandtl number  (\ref{Pm}).   The  magnetic Reynolds number
\begin{equation}  \label{defin-Rm}
Rm = Pm \cdot Re 
\end{equation}
is also generally used.  Unlike $Pm$,  it depends on the choice of $L$ in the definition of the Reynolds number.

We are interested in the stability condition $Pm_c(Re_{\lambda})$, or $Rm_c(Re)$  such that there is a solution of (\ref{eq-psi})  with $\gamma = 0$  and there is no solution with $\gamma>0$ for smaller  $Rm$.   We note that the second relation is conventional, but the first one is more universally defined. 

In the vicinity of this stability curve, $\gamma$ is known to depend log-linearly on $Rm$~\citep{rogachevskii1997intermittency, kleeorin2012growth}, \begin{equation*}\gamma \propto \ln (Rm/Rm_c).\end{equation*}  Based on Eq. (\ref{eq-psi}), we calculated the proportionality coefficient numerically and compared it with the results of \citet{warnecke2023numerical}.   
To solve Eq. (\ref{eq-psi}), we assumed some model for $b(\rho)$.  This produced one more difficulty, since $b(\rho)$ is the quasi-Lagrangian correlation function and is difficult to measure~\citep{biferale2011multi}. By definition of the  Lagrangian correlation time $\tau_c(\rho)$ introduced by~\citet{l1997temporal}, we have
\begin{equation}     \label{b-ot-rho}
b(\rho) =  \ls (\delta v_{\parallel} )^2\rs (\rho) \, \tau_c (\rho) 
.\end{equation}
In~\citet{biferale2011multi,donzis2010bottleneck}, the  simultaneous structure function $ \ls (\delta v_{\parallel} )^2\rs $ and the correlation time $\tau_c $  were found  from DNS for $Re_{\lambda}=140$;  for much higher $Re_{\lambda}$, we used theoretical considerations~\citep{l1997temporal} that argue that both $ \ls (\delta v_{\parallel} )^2\rs $ and $\tau_c $  are power-law functions of $\rho$ inside the inertial range. 

\section{$Re_{\lambda} = 140 $: Moderate $Re$ and $Pm$ }~\label{S:3-140}
In this section, we refer to the results of~\citet{biferale2011multi,donzis2010bottleneck} for data on velocity statistics and~\citet{iskakov2007numerical,schekochihin2007fluctuation} for data on magnetic generation. Fortunately, all these papers contained the DNS performed for  the same Reynolds number $Re_{\lambda}=140$.  
The critical Prandtl number is not small for this Reynolds number, so the bottleneck region and even the viscous range of scales can affect the results and have to be taken into account. The main parameters of the DNS performed by~\citet{biferale2011multi} are listed in Table~\ref{tab:1-bif_parameters}.

\begin{table}
\centering
\caption{Main parameters of the flow with $\mathrm{Re}_{\lambda}=140.$}
\label{tab:1-bif_parameters}
\begin{tabular}{cl}
\toprule
Parameter &  Value \\
\midrule
Integral scale & $L=4.24$ \\
Taylor scale & $\lambda=0.30$ \\
Kolmogorov viscous scale & $r_\nu=1.28 \times 10^{-2}$ \\
Energy flux & $\varepsilon=1$ \\
Kinematic viscosity & $\nu=3 \times 10^{-3}$ \\
Single-component RMS velocity &  $u'=1.41$ \\
\bottomrule
\end{tabular}
\tablebib{
\citet{biferale2011multi}.
}
\end{table}

\citet{biferale2011multi} calculated the dependence of the Lagrangian correlation time $\tau_c$ on $\rho$, 
\begin{align}\label{tau-bif}
&\tau_c   
(\rho)=C_1\,t_\nu\frac{\left(1+\left( C_2  \rho/r_\nu\right)^2\right)^{1/3}}{\left(1+\left( C_3  \rho/L\right)^2\right)^{1/3}},
\end{align}
where
\begin{equation*}
C_1=2.15 \ , \quad   C_2=6.75\cdot 10^{-2}  \ , \quad  C_3=2.94  \ , 
\end{equation*}
$r_{\nu} = \nu^{3/4} \varepsilon^{-1/4}$  is the Kolmogorov viscous scale,    and $t_{\nu} = \sqrt{\nu/\varepsilon}$ is the viscous characteristic time.

The second-order simultaneous  velocity structure function is well investigated; in the viscous and inertial ranges, its behavior is universal and is very well described by the generalized Batchelor approximation~\citep{donzis2010bottleneck},
\begin{align}\label{v2}
\ls (\de v_{\parallel})^2\rs=\frac{\varepsilon}{15\nu}\frac{\rho^2}{\Bigl(1+\left(C_B\,\rho/r_\nu\right)^q\Bigr)^{\frac{2-\zeta_2}{q}}},\,\,\,\,\,\,\,\,\,\rho\ll L,
\end{align}
\begin{equation*}   C_B=7.6\times10^{-2} \ , \quad   \zeta_2=0.67 \ , \quad  q=1.82  .\end{equation*}
It is easy to see that in the ultraviolet regime, it coincides with Kolmogorov's analytical formula, 
$\ls (\de v_{\parallel})^2\rs = \frac{\varepsilon}{15\nu} {\rho^2}$, $\rho \ll r_\nu$,  while    inside  the inertial range,  it demonstrates a power-law behavior,  
$\ls (\de v_{\parallel})^2\rs  \propto   {\rho^{\zeta_2}}$, $L \gg \rho \gg r_\nu/C_B$.  We note that the coefficients $C_B$  in  (\ref{v2}) and $C_2$ in 
(\ref{tau-bif})  are both comparably  small. This corresponds to the existence of the bottleneck transition region between the viscous and the inertial ranges: viscosity is essential at scales $\rho \sim r_{\nu}/C_2 \simeq r_{\nu}/C_B$, which are significantly larger than the viscous scale.

For our purpose, we enlarged the limits of applicability of  (\ref{v2}) for larger scales up to the integral scale,  where $\ls (\de v_{\parallel})^2\rs = 2 u'^2 $, $\rho \geq   L$. In accordance with the idea of extended self-similarity (ESS), this can also be done by means of a fractionally rational approximation \footnote{For this shape of the correlation functions,    
\begin{equation*}   \frac{\zeta_2(r)}{\zeta_3(r)} \equiv \frac{d \ln \langle (\delta v_{\parallel})^2\rangle}{d \ln \langle |\delta v_{\parallel}|^3\rangle} = const
\end{equation*} in accordance with ESS~\citep{benzi1993extended}. },

\begin{align}\label{v2-main}
&\ls(\de v_{\parallel} )^2\rs _{({\rm main})}
=\frac{\varepsilon}{15\nu}\frac{\rho^2}{\Bigl(1+\left(C_B\,\rho/r_\nu\right)^q\Bigr)^{\frac{2-\zeta_2}{q}}\Bigl(1+\left(C_m\rho/L\right)^2\Bigr)^{\frac{\zeta_2}{2}}}, 
\end{align}
where $C_m=1.6$  follows from the matching with the integral scale.  

\begin{table}
\centering
\caption{Results of the Kazantsev equation analysis and DNS ($\mathrm{Re}_{\lambda}=140$).}
\label{tab:2-140}
\begin{tabular}{lccc}
\toprule
$Re_\lambda=140$ & Approx. & $Pm_c$ & ${Rm}_c^{\rm Sch}$ \\
\midrule
Main model & (\ref{tau-bif}), (\ref{v2-main}) & 0.26 & 170 \\
Sharp sub-integral transition & (\ref{tau-bif}), (\ref{v2-sharp}) & 0.22 & 140 \\
Sharp bottleneck & (\ref{tau-bif}), (\ref{v2-sharp-bottleneck}) & $0.29$ & 190 \\
DNS\tablefootmark{a}  & -- & 0.3 & 195 \\
\bottomrule
\end{tabular}
\tablefoot{
Critical magnetic Prandtl numbers and magnetic Reynolds numbers at $\mathrm{Re}_{\lambda}=140$ for different theoretical models and DNS.
\\
\tablefootmark{a} \citet{schekochihin2007fluctuation}.
}
\end{table}

The two transitional regions,  the bottleneck and the range between the inertial and integral scales,  are presented in (\ref {v2-main}) by the two brackets in the denominator.   To estimate the accuracy of the approximation and to visualize the contributions of the two  transition regions, we consecutively substituted  a piecewise power-law function for each of the brackets, 
\begin{multline} \label{v2-sharp}
\ls(\de v_{\parallel} )^2 \rs_{({\rm Sharp-int})}=\frac{\varepsilon}{15\nu}\frac{\rho^2}{\Bigl(1+\left(C_B\,\rho/r_\nu\right)^q\Bigr)^{\frac{2-\zeta_2}{q}}}\,\theta(  L/a_{\rm int} - \rho)+
\\2\,u'\,\theta(\rho - L/a_{ \mathrm{int}}) 
\end{multline} 
and 
\begin{multline}\label{v2-sharp-bottleneck}
\ls(\de v_{\parallel} )^2\rs_{({\rm Sharp-bot})}=\frac{\varepsilon}{15\nu}   \frac{\rho^2}
{\Bigl(1+\left(C_m\rho/L\right)^2\Bigr)^{\frac{\zeta_2}{2}}}      \,\theta( \rho -  a_{\rm bot
}r_\nu )+ 
\\
\frac{\varepsilon}{15\nu}   \rho^2  \,\theta( a_{\rm bot}r_\nu - \rho)    \   . 
\end{multline} 
The coefficients $a_{\rm bot} = 13.1  $  and  $a_{\rm int} = 1.6 $ were also found from the matching condition, as were $C_B$ and $C_m$.       The  value of $a_{int}$ coincides to high accuracy with $C_m$, which indicates that the two approximations are rather close. 

The results are presented in Table~\ref{tab:2-140}. 
The theoretical predictions are rather close to the numerical results; the use of the Eulerian structure function $b(\rho)$  would give an estimate of $Rm_c$ at least an order of magnitude lower.
The fact that the theoretical prediction for $Rm_c$ is lower than the experimental result may be caused by the assumption of  Gaussianity of the flow: when the declination from Gaussianity is taken into account, that is, the third-order correlator of the velocity, the result can increase~\citep{kopyev2024suppression}.
 
\begin{figure}
\centering
\resizebox{7cm}{!}{\includegraphics{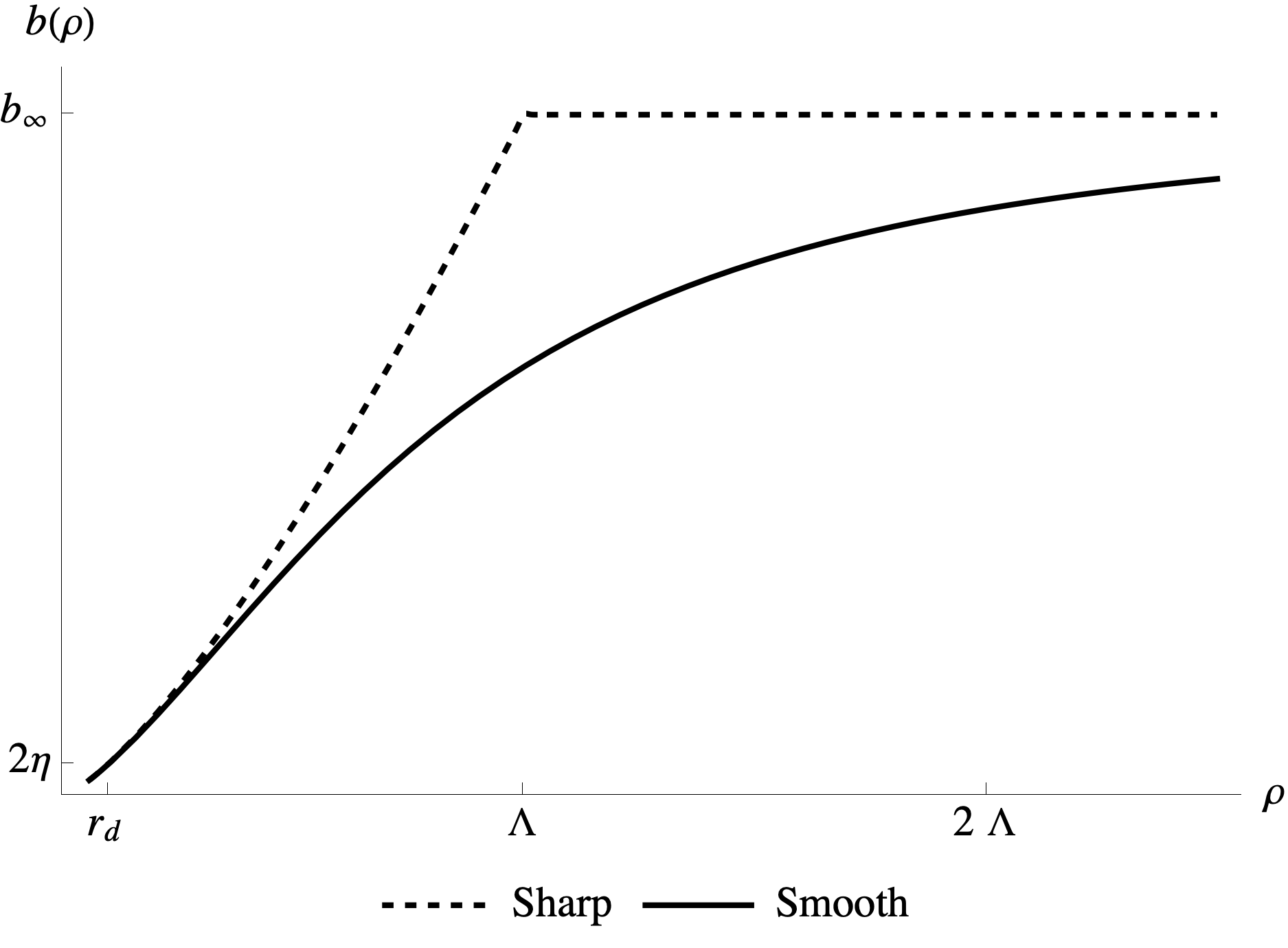}}
\caption{Shape of $b(\rho)$ for the Sharp (Eq.~\ref{b-Sharp}) and Smooth (Eq.~\ref{b-Smooth}) models for the same values of the parameters $s$, $b_{\infty}$, and $\Lambda$.}
\label{fig:1-bmodels}
\end{figure}
The critical values of  the magnetic Reynolds number and Prandtl number are close for all the three models, but the results still differ by about 15\%.  This indicates  the effect of the transitional regions on the near-threshold generation.  
The importance of the bottleneck region is determined by the  fact that the  magnetic    Prandtl number is not small enough to neglect the effect of the viscous range. From the value of $C_B$ in (\ref{v2-main}) and $C_2 $ in (\ref{tau-bif}),  it follows that the effect of the viscosity reaches about $15 r_{\nu}$, while for $Pm = 0.25$, the magnetic diffusivity scale is only about $3 r_{\nu}$, so it lies deep inside the bottleneck range.  It is natural to assume that for smaller $Pm$, the effect of  viscosity on  the magnetic field generation decreases. 
In contrast, the effect of  the external transition  between the inertial and the integral ranges   can only  become even stronger for higher Reynolds numbers. 

\section{Very high Reynolds numbers}~\label{S:4-largeRe}

\noindent The lack of experimental or DNS data on the hydrodynamic Lagrangian correlation time prevented us from determining $b(\rho)$ for other Reynolds numbers, and, hence, from calculating $Rm_c$ for them.
However, for a well-developed turbulence with a wide inertial range, we assumed in accordance with the general theoretical approach to turbulence that $b(\rho)$ has a universal shape regardless of the details of the flow.  To normalize the function, we took the largest scales. For these scales,  
\begin{equation}   \label{deltav^2}
\lim_{\rho\to\infty}\ls (\delta v_{\parallel})^2\rs=\frac{2}{3}v_{\rm rms}^2= 2{u'}^{\,2},
\end{equation}
and the integral Lagrangian correlation time   $T_L=\lim \limits_{\rho \to \infty} \tau_c$  can be found from the approximation based on Sawford’s second-order stochastic model~\citep{sawford1991reynolds, sawford2008reynolds},
\begin{equation}\label{T-L-as}
T_L=\sqrt{\frac{\nu}{\varepsilon}}\frac{2Re_\lambda}{\sqrt{15}C_0} = \frac{2}{C_0} \frac{u'^2}{\varepsilon},
\end{equation}
where $C_0$ is the Kolmogorov constant. It is rather difficult to find~\citep{lien2002kolmogorov, ouellette2006small, zimmermann2010lagrangian, uma2025numerical}, and we used the value obtained  by~\citet{sawford2011kolmogorov},
\begin{equation}\label{C0}
C_0=6.9\pm0.2.
\end{equation}

In what follows, we therefore consider different models for $b(\rho)$  that  satisfy the normalization condition consistent with (\ref{b-ot-rho})
\begin{equation}  \label{b-infty}
 b_{\infty} \equiv    b ({\infty} )  =   2 u'^2 T_L .
\end{equation}
According to 
(\ref{T-L-as}),   (\ref{Re-lambda}),(\ref{Pm}), 
\begin{equation}  \label{intermed}
\frac{b_{\infty}}{2\eta} =  Pm \frac{Re_{\lambda}^2}{15} \frac{2}{C_0} \propto  Rm .
\end{equation}
In what follows, we use the dimensionless value 
\begin{equation}  \label{X-definit}
X= \left( \frac{b_{\infty}}{2\eta} \right) ^{1/(1+s)} 
\end{equation}
to characterize the generation properties  of a flow.  Here, $s$ is the scaling exponent of    $\langle |\delta v_{\parallel}|\rangle$. 
The value of $X$ describes the relative width of the range of scales that  can affect the generation.  We note that for any particular series of DNS or experiments,   $Rm \propto X^{1+s}$, but  because the definition of the integral scale $L$ is uncertain, $X$ is more accurately  defined than $Rm$. 

\subsection{Sharp model} 
We started with the natural and simple theoretical model proposed by~\citet{vainshtein1986dynamics}, which focuses on the main features of turbulence (hereinafter, model 'Sharp'). It describes the power-law behavior of $b(\rho)$  in the inertial range and its approximate  constancy at large scales,
\begin{equation}\label{b-Sharp}
b(\rho)_{\rm Sharp} = b_{\infty} \left\{ \begin{array}{ll}   
(\rho/\Lambda)^{1+s}  \ \ ,  \ &  \rho <\Lambda \\
1 \ ,& \rho > \Lambda   \end{array} \right.  \ ,   \qquad  s=1/3    .
\end{equation}
This model was used in numerous papers (see, e.g., \citealt{rogachevskii1997intermittency, vincenzi2002kraichnan, boldyrev2004magnetic, arponen2007dynamo, schober2012small, kleeorin2012growth, kopyev2025virtual}).  The viscous range is not of interest for a dynamo study at very high Reynolds numbers, since  it is far  below the magnetic diffusion scale and cannot contribute to the generation.   The integral scale $L$  used in the definition of $Re, Rm$  was assumed to be proportional to $\Lambda$.
The sharp break in Eq. (\ref{b-Sharp})   results in a kink in $\sigma(\rho)$ and in a discontinuity in $\sigma'$. This discontinuity produces a $\delta$-function  in the potential (\ref{U-pot}). 
This  is a peculiar property of the model  and just enhances and emphasizes the maximum of $U(\rho)$,  but  the term $\rho \sigma'$   does not play an essential role. It changes the resulting $Rm_c$ to only about 20\%~\citep{kopyev2025virtual}.
A much larger  contribution to  $Rm_c$ is produced by  the sharp kink at the same $\rho =\Lambda$.
\begin{table}
\centering
\caption{Results for very high Reynolds numbers.}
\label{tab:3-high_Re_comparison}
\begin{tabular}{ccccc}
\toprule
& \multicolumn{2}{c}{Sharp} & \multicolumn{2}{c}{Smooth} \\
\cmidrule(lr){2-3}\cmidrule(lr){4-5}
$s$ & 0.33 & 0.39 & 0.33 & 0.39 \\
$X_c$ & 20.5 & 14.5 & 43.2 & 30.2 \\
${Rm}_c^{\rm Sch}$ & $95 \pm 5$ & $70 \pm 5$ & $260 \pm 10$ & \bfseries $200 \pm 10$ \\
\midrule
\multicolumn{5}{l}{DNS\tablefootmark{a, b}:} \\
\multicolumn{5}{l}{\quad ${{Rm}^{\rm Sch}_c}_{\max} \approx 350$ at ${Re}_{\lambda} =300$} \\
\multicolumn{5}{l}{\quad ${Rm}_c \downarrow $ for higher ${Re}$} \\
\multicolumn{5}{l}{\quad ${Rm}_c^{\rm Sch} ({Re}_{\lambda}\simeq 5 \times 10^4) \gtrsim  200$} \\
\bottomrule
\end{tabular}
\tablefoot{Comparison of ${Rm}_c^{\rm Sch}$ for different models and DNS. The uncertainties of ${Rm}_c$ are determined by the uncertainty of $C_0$~(Eq.~\ref{C0}).
\\
\tablefootmark{a} \citet{schekochihin2007fluctuation}, \tablefootmark{b} \citet{warnecke2023numerical}.}
\end{table}

From (\ref{X-definit}), $X$ is the ratio of the end of the inertial range scale $\Lambda$ and the magnetic diffusion scale $r_d$,
\begin{equation*}
X   = \left( \frac{b_{\infty}}{2\eta} \right) ^{1/(1+s)}  =   \frac{\Lambda}{r_d}  \qquad \mbox{where}  \quad  r_d : \ b(r_d) = 2\eta.
\end{equation*}

\citet{kopyev2025virtual} reported that the generation threshold corresponds to a critical value of $X$ equal to
\begin{equation*}
X_c = 20.5.
\end{equation*}
Then, according to (\ref{intermed}) and (\ref{Re-Schek}),
\begin{equation*}
{Rm_c}^{\rm Sch}_{\rm(Sharp)} = \frac{C_0}4 X_c^{4/3} \simeq 100  .
\end{equation*}
In what follows, we compare this result with the results of other models.

\subsection{Intermittency: Dependence on $s$}
The scaling exponent of $b$  inside the inertial range is an important parameter.  It is composed of the exponents of $\langle (\delta v_{\parallel})^2\rangle$ and $\tau_c$.   Let  $\langle |\delta v_{\parallel}|^n\rangle \propto  \rho^{\zeta_n}$ ; then, according to the bridge relations \citep{l1997temporal, biferale2011multi}, $\tau_c \propto \rho ^{1-\zeta_2+\zeta_1}$  and, hence,
\begin{equation}
b(\rho) \propto \rho^{1+\zeta_1}.
\end{equation}
In the Kolmogorov phenomenological theory, we obtain $s= \zeta_1 = 1/3$.  However, the intermittency of the velocity structure functions implies that    $\zeta_n>n/3$  for $n<3$  \citep{frisch1995turbulence}.  

The values of  $\zeta_n$  depend on $Re$; for very high Reynolds numbers ($Re_\lambda\gtrsim650$), we obtain $s=\zeta_1 = 0.39 $~\citep{benzi2010inertial, iyer2020scaling}.   
Even this small change in $s$  affects the generation properties significantly: for $s=0.39$, we obtain $X_c = 14.6$, which results in  smaller critical Reynolds number,
\begin{equation*}
Pm_c  \simeq  \frac{2160}{Re_{\lambda}^2}  \ , \quad Rm_c^{\rm Sch} \simeq 70.
\end{equation*}
This qualitative behavior is an important indication for the explaination of the decrease in $Rm_c$ for high enough Reynolds numbers. \citet{iyer2020scaling} showed that 
$\zeta_1$ increases as a function of $Re$. This is a sufficient reason for the decrease in $Rm_c$ observed by~\citet{warnecke2023numerical}.
 
\subsection{Smooth model: Effect of the transition region}

\begin{figure}
\centering
\includegraphics[width=\linewidth]{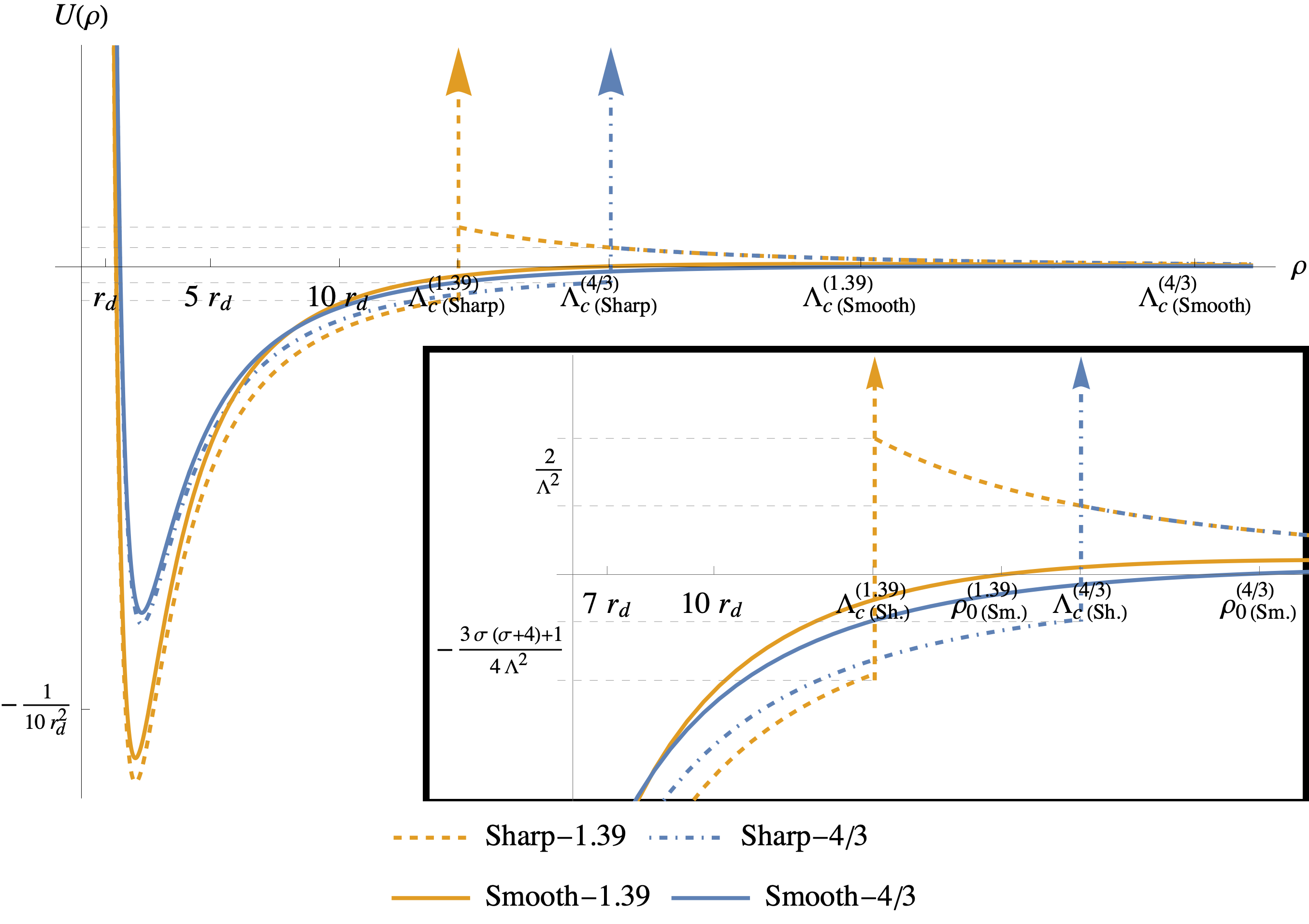}
\caption{Effective potential $U(\rho)$ (Eq.~\ref{U-pot}) that corresponds to the generation threshold for the two Sharp and two Smooth models. The length scale is normalized by the diffusion scale $r_d$, which is taken to be the same for both models. The vertical arrows correspond to the $\delta$ functions in the Sharp model potential.}
\label{fig:3-PotentialModels}
\end{figure}

Not only the inertial range, but also the transition scales from the inertial to the integral range affect the generation of high Reynolds numbers. To  demonstrate this, we considered the 'Smooth' model: it coincides with the Sharp model in the inertial and integral ranges, but provides a more accurate description in between,
\begin{equation}  \label{b-Smooth}
b(\rho)_{{\rm Smooth}} = b_{\infty}   \frac{(\rho/\Lambda)^{1+s} }{\left(  1+(\rho/\Lambda)^2  \right) ^\frac{1+s}2} .
\end{equation}
Since at $\rho  \ll \Lambda$  and at $\rho \to \infty$  this coincides with $b_{\rm Sharp}$,  the definition of  $r_d$  and the relation $X=\Lambda/r_d$ remain the same.   The shape of $b(\rho)$ for the Smooth and Sharp models is illustrated in Fig.~\ref{fig:1-bmodels}.

We considered the Smooth model with a classical nonintermittent  scaling $s=1/3$ and with a more realistic $s=0.39$.  The resulting $X_c$ and $Rm_c^{\rm Sch}$ are presented in Table~\ref{tab:3-high_Re_comparison}.
 Comparing the results of the theoretical models with the DNS data, we found that the agreement is quite reasonable. All the models show a value of  $Rm_c$ lower than the upper limit $Rm_c^{\rm Sch} \le 350$ found by~\citet{schekochihin2007fluctuation}.   The data of  \citet{schekochihin2007fluctuation,iskakov2007numerical,warnecke2023numerical}  demonstrate  the decrease in $Rm_c$ at  $Re_{\lambda} \gtrsim 300$, and  the value of $Rm_c$ obtained at the highest  Reynolds number  $Re\simeq 5 \cdot 10^4$  achieved by~\citet{warnecke2023numerical}\footnote{The paper \citet{warnecke2023numerical}  reports   $Rm_c \approx 10^2$ , but the normalization of $Re$ and $Rm$  that is used there  
is about two times lower than that of \citet{schekochihin2007fluctuation, iskakov2007numerical}: according to \citet{brandenburg2018varying}, the definition of $Re$  (and, hence, $Rm$) in \citet{brandenburg2018varying} is probably $1.5$ times smaller than those in \citet{iskakov2007numerical} because of different definitions of the pumping scale.  On the other hand, the values of $Rm_c$ corresponding to $Pm=0.1$ in \citet{brandenburg2018varying} and \citet{warnecke2023numerical}  , for example, are related as 4:3. So,  $$Rm^{\rm Sch} = Rm^{\rm Isk} \approx \frac 32  Rm^{\rm Brand}  \approx 2 Rm^{\rm W}$$.     }   is $Rm_c^{\rm Sch} \gtrsim 200$. This agrees very well with the prediction of the Smooth model for $s=0.39$, which is the most realistic of our models.  

  \begin{figure}
    \includegraphics[width=\linewidth]{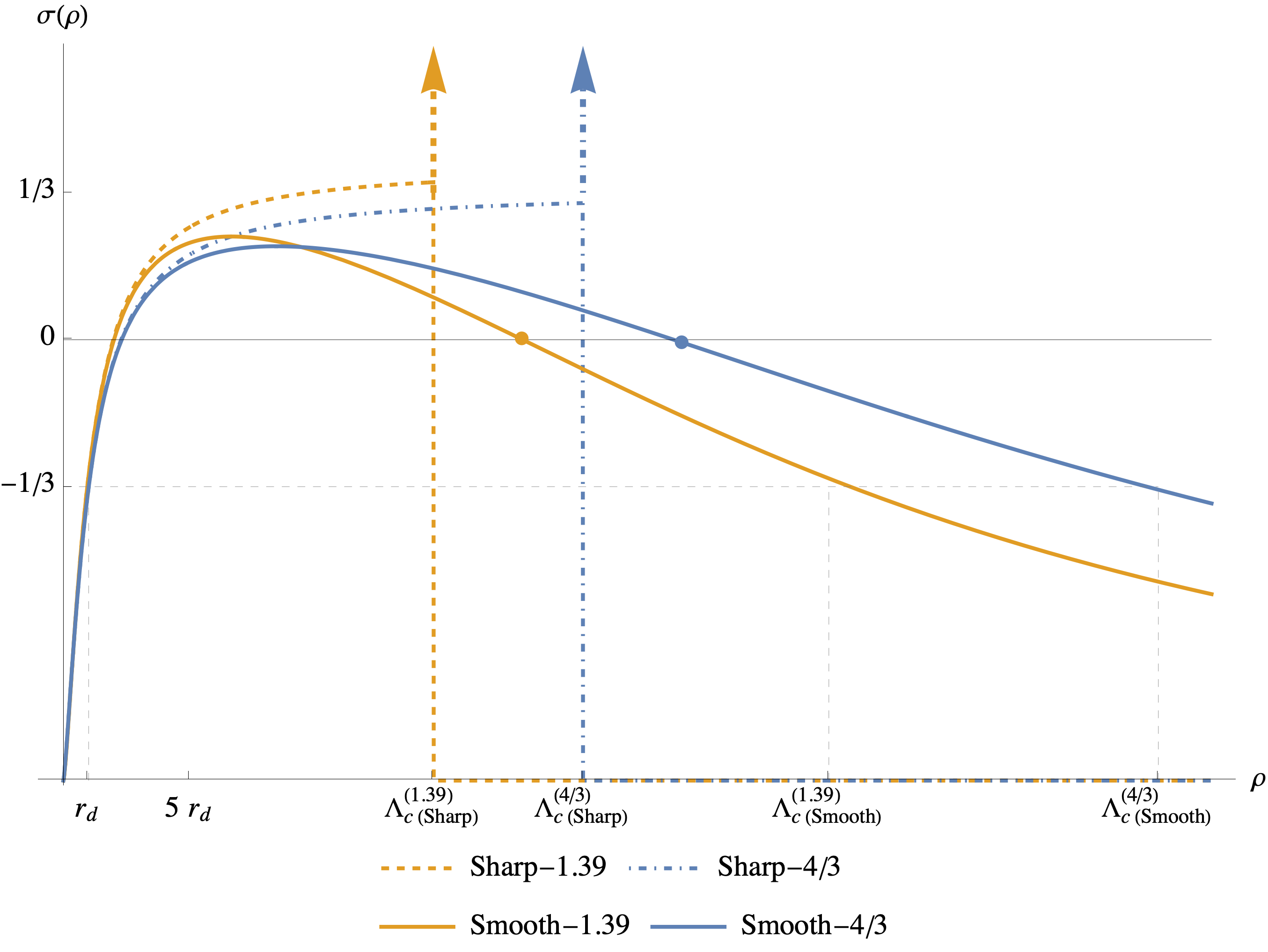}
 \caption{Logarithmic derivative $\sigma (\rho)$,  (\ref{sigma}), for the four models. The length scale is normalized by the same $r_d$.  The parameters of each model correspond to the generation threshold. } 
 \label{fig:4-LogDerivModels}
\end{figure}

Comparison of the Smooth and Sharp models shows  
that the effect of the transition region
is significant: the critical values of the parameters differ by almost a factor of two. 
 We determine the reason of  the  difference below.
 
The velocity correlator profiles and the corresponding potentials $U(\rho)$  drawn at the critical values $\Lambda_c$  are rather close in both models everywhere except in the vicinity of  $\Lambda_{c ({\rm Sharp})}$ (see Fig.~\ref{fig:3-PotentialModels}).  
This appears to be the region in which the transition zone has the strongest effect.  
On the other hand, this is the boundary of the region in which the generation occurs. The generation threshold  at any given Reynolds number corresponds to the zero energy level; 
Fig~\ref{fig:3-PotentialModels} shows that  the scales $\rho_{0} \, : \ U(\rho_0)=0$ for the Smooth models differ by less than 20\% from the corresponding scales $\rho_{0 ({\rm Sharp})} = \Lambda_{c ({\rm Sharp})}$ for the Sharp models. 
Hence, the generation region $\rho \lesssim \rho_0$  is situated for the Sharp and Smooth models  within  or near the scale $\Lambda_{c ({\rm Sharp})}$  and rather far from the integral scale $\Lambda_{c ({\rm Smooth})}$.  Although the Sharp model is less plausible and accurate than the Smooth model,  
it indicates the scales that contribute most to the generation better. 

The difference in  the values of $Rm_c$  is therefore probably caused by different  meanings   of the scale  $\Lambda$  in the two types of models.  While $\Lambda_{c ({\rm Smooth})}$ corresponds to the integral scale of turbulence,  $\Lambda_{c ({\rm Sharp})}$  merely acts as a transition range marker, which is about twice smaller.  

The function $b(\rho)$ contributes to $U(\rho)$  by means of the logarithmic derivative $\sigma(\rho)$. It is therefore interesting to compare these functions in critical regimes for different models.  
In Fig.~\ref{fig:4-LogDerivModels} the functions $\sigma(\rho)$ that correspond to the generation threshold are plotted. The rising parts of the curves are almost identical,  the whole difference being in the descending part of the curve. 
In accordance with~\citet{kazantsev1968enhancement}, the scales important for the generation satisfy the condition $\sigma >0$. For every $s$,  the graphs for both models intersect the line $\sigma =0$  
at  $\rho  \simeq {\Lambda_c}_{({\rm Sharp})}$. 
This shows that the Kazantsev criterion agrees well with the criterion $\rho \lesssim \rho_0$. 

The transition region probably determines the 
generation  boundary. 
On the other hand, in the absence of the self-similarity of velocity correlators at the largest scales,
the properties of this transition region may not be uniquely related to the velocity properties at the integral scales.  
If this is so, the comparison of some characteristics related to the transition region, for example, $\rho_0$ or the scale at which  $\sigma =0$ for different DNS-s or experiments, would be more effective than the comparison of large-scale parameters such as the magnetic Reynolds numbers.

In other words, it may be more effective to describe the generation threshold by means of some combination of parameters that describe the surrounding of the point  $\rho_0$ or  $\rho: \sigma =0$ rather than  by means of integral characteristics such as $Rm_c$.

\section{Increment near the generation threshold}~\label{S:5-incr}

We considered the growth rate  $\gamma = \lim \limits_{t\to \infty }  d \ln G/dt $  of the magnetic field correlator near the generation threshold. Namely, we were interested in 
\begin{equation*}
g \equiv \left.  Rm \frac{d \gamma }{d Rm } \right|_{Rm=Rm_c} =    \left. \frac{d \gamma }{d \ln (Rm/Rm_c) } \right|_{Rm=Rm_c}.
\end{equation*}
We note that since the expression contains the ratio $Rm/Rm_c$ , it does not depend on the choice of normalization. 

\citet{rogachevskii1997intermittency, kleeorin2012growth} predicted this value to be constant along the critical curve ($Rm=Rm_c$) for small Prandtl numbers.  The data of~\citet{warnecke2023numerical}   confirmed this prediction for a wide range of Prandtl numbers beginning  with $Pm \lesssim 0.1$. The value of the coefficient found by~\citet{brandenburg2018varying, warnecke2023numerical} is
\begin{equation*}
\left( \frac{\gamma_{\tiny\mathrm{DNS-W}}}{ \ln (Rm/Rm_c) }\right)_{Pm}= 0.022 v_{{\rm rms}}  k_f ,
\end{equation*}
where $k_f$ is the pumping wave number;  taking the relation 
$  {\varepsilon}/({v_{\rm rms}^3 k_f}) = 0.041 $  given by~\citet{brandenburg2018varying} into account, we can express this in terms of universal variables,
\begin{equation}   \label{increment-DNS}
\left( \frac{\gamma_{\tiny\mathrm{DNS-W}} }{ \ln (Rm/Rm_c) } \right) _{Pm}   = 0.18 \frac{\varepsilon}{u'^2}.
\end{equation}

To compare this coefficient   with the Kazantsev theory, we were again forced to take 
 the value $Re_\lambda = 140$, since for this Reynolds number, we have information on the Lagrangian statistics~\citep{biferale2011multi}.   The Prandtl number corresponding to this $Re_{\lambda}$ is $Pm  \approx 0.3 $ and does not  belong to the range in which the increment  slope is constant.   However,  according to the data  of~\citet{warnecke2023numerical} ,    the value  is close to the boundary and the slope does not differ significantly. 
 
Making use of the velocity statistics  given by~\citet{biferale2011multi,donzis2010bottleneck}, we solved Eq. (\ref{eq-psi})  numerically for values of $\gamma$ close to zero and obtained (for $Re_{\lambda}=140$)
\begin{equation}   \label{something}
\left(  \frac{\gamma_{\rm th} }{ \ln (Rm/Rm_c) }\right)_{Re}  \simeq  1.3  \, \varepsilon / u'^2.
\end{equation} 
To compare this with Eq. (\ref{increment-DNS}) , we note that the increment derivatives in  (\ref{increment-DNS})  are taken at  constant Prandtl numbers, while  in Eq. (\ref{something}), they are taken at  constant Reynolds numbers.  These two derivatives are  related by
\begin{equation*}
\frac{ (\partial  \gamma / \partial \ln Rm)_{Pm}}{ (\partial  \gamma / \partial \ln Rm)_{Re}} = 
-\frac{Re}{Pm_c} \frac{dPm_c}{d Re}  = - \frac{d \ln Pm_c}{d \ln Re}.
\end{equation*}
This follows from Eq. (\ref{defin-Rm})   and from the condition that  $\gamma=0$  along the critical curve $Rm_c(Re)$.
From the data presented by~\citet{schekochihin2007fluctuation,warnecke2023numerical}, we found
\begin{equation*}
- \frac{d \ln Pm_c}{d \ln Re} = \left(  1- \frac{d \ln Rm_c}{d \ln Pm} \right)^{-1}  \simeq 0.5.
 \end{equation*}
\begin{table}
\centering
\caption{Simulation data of \citet{schekochihin2007fluctuation}}
\label{tab:4-Sch-DNS}
\setlength{\tabcolsep}{6pt}
\begin{tabular}{cccccccc}
\toprule
Run & ${Rm^{\rm Sch}}$ & ${Re^{\rm Sch}}$ & $Pm$ & ${Re_\lambda}$ & $\gamma$ & $u'$ & $\varepsilon$ \\
\midrule
A3 & 110 & 440 & 0.25 & 111 & $-$0.22 & 0.8 & 1 \\
B2 & 220 & 440 & 0.5 & 110 & 0.49 & 0.8 & 1 \\
\bottomrule
\end{tabular}
\tablefoot{Results of two simulations. We used them to obtain~\eqref{increment-DNS-Sch}.}
\end{table}

The results for the theoretical calculation and for the DNS results are summarized in Table~\ref{tab:5-gamma-compar}. We also added $g$ obtained from  the  DNS data~\citep{schekochihin2007fluctuation} (see Table~\ref{tab:4-Sch-DNS}), 
\begin{equation}\label{increment-DNS-Sch}
\left(   \frac{\gamma_{\tiny\mathrm{DNS-W}}}{ \ln (Rm/Rm_c) }\right)_{Re}\simeq\frac{0.49+0.22}{\ln(220/110)} \simeq 0.7 \frac{\varepsilon}{u'^2}.
\end{equation} 

The results differ significantly.    Even the results of the two numerical simulations are quite different;  when this is taken into account, the fact that the theoretical and numerical results  are on  the same order  is a good correspondence.   

The significant difference between the theoretical and numerical results can be explained as follows, for example: \\
- The data for hydrodynamics are taken from different sources, and this affects the errors strongly.  In particular, $C_0$ is difficult to determine and depends strongly on the isotropy of the flow and other factors. We used the value (\ref{C0}) found analytically~\citep{sawford1991reynolds} in the frame of a particular model.  The difference given by various  DNSs and experiments is  up to 1.5 times in  both directions~\citep{lien2002kolmogorov, uma2025numerical}.  \\
- In Table~\ref{tab:4-Sch-DNS}
the estimate of the  slope of  a curve near the inflection point  is taken  by only two distant points, which may lead to an underestimation of the slope. In addition, these two points are themselves known to have some errors.  \\
- The details of DNS, such as periodic boundary conditions, can affect the generation. \\
- Finally,
the real slope may be smaller than the one predicted by the Kazantsev model because of the non-Gaussianity of the velocity flow. If this factor were taken into account,  the slope would decrease \citep{kopyev2024suppression}.

\begin{table}
\centering
\caption{Results for the increment slope}
\label{tab:5-gamma-compar}
\begin{tabular}{lccc}
\toprule
& ${Re}_\lambda$ & $g_{Pm}$ & $g_{Re}$ \\
\midrule
Theory & $140$ & $0.66$ & $1.33$ \\
DNS\tablefootmark{a} & $110$ & $0.35$ & $0.7$ \\
DNS\tablefootmark{b} & $\gtrsim 100$ & $0.2$ 
& 0.36 \\
\bottomrule
\end{tabular}
\tablefoot{
The increment slope $d {\gamma }/d{ \ln (Rm/Rm_c) }$ for $Pm$=const ($g_{Pm}$) and $Re$=const~($g_{Re}$) obtained from the theory and the DNS.
\\
\tablefootmark{a} \citet{schekochihin2007fluctuation}, \tablefootmark{b} \citet{warnecke2023numerical}.
}
\end{table}

 \vspace{0.2cm}

In the case of high Reynolds numbers, we lack enough data for nonsimultaneous velocity statistics, and we restricted ourselves by dimensional considerations.  For the Sharp and Smooth models, 
we obtained inside the inertial range of scales 
\begin{equation} \label{b-stepen}
\left. {b(\rho)} \right|_{r_\nu \ll \rho \ll L} = \rho^{4/3} \left( \frac{\rho}{\Lambda}   \right)^{s-1/3}  \varepsilon^{1/3} \tilde{a} (Re_{\lambda}) \ ,
\end{equation}
where $\tilde{a}$  is a dimensionless parameter.   As $Re\to \infty$, the function $\tilde{a}(Re_{\lambda})$ has a constant limit~\citep{frisch1995turbulence}.
Taking (\ref{b-infty}) into account, we obtain for the Sharp model from the condition $b(\Lambda)= b_{\infty} $
\begin{equation*}
\Lambda = \left(  \frac{b_{\infty}}{\varepsilon^{1/3} \tilde{a}}\right) ^{3/4} = \left(  \frac{4}{C_0 \tilde{a}}\right)^{3/4}\frac{u'^3}{\varepsilon   } .
\end{equation*}
For the Smooth model, the result differs no more than by a multiplier of $\sim 2$. 
Then, the increment is  
\begin{equation*}
g   =  \frac{c X_c^{s-1}}{1+s} \frac{2\eta}{r_d^2} = 
 \frac{c }{1+s}\frac {b_{\infty}}{\Lambda^2} = \frac{c }{1+s}  \sqrt{\frac{C_0}4} \frac{\varepsilon}{u'^2} \tilde{a}^{3/2} .
\end{equation*}
This confirms  that the increment does not depend on the Reynolds number for very high $Re$.  \citet{kleeorin2012growth} obtained  this result for $s=1/3$;  we prove that the presence of  intermittency  practically does not affect it, so that the independence of $g$  is preserved up to the values of $Re$ at which the bottleneck effect is essential. 

\section{Discussion}

We studied the magnetic field generation in a turbulent flow near the generation threshold. We compared  theoretical predictions of the extended Kazantsev theory with  results of DNS   for two cases: the value of the local Reynolds number $Re_{\lambda}=140$, which is distinguished by the presence of magnetohydrodynamical and Lagrangian hydrodynamical  data, and the limiting case of very high Reynolds numbers.  To calculate the time-integrated velocity structure function $b(\rho)$ (see Eq.~\ref{b-def}),  we used the quasi-Lagrangian statistics (in particular, the Lagrangian velocity correlation time).  
This provided a  better concordance with the DNS results by some orders of magnitude than  those found with  the
Eulerian statistics. 
The values of critical magnetic Reynolds and magnetic Prandtl numbers  obtained from the theory and from DNS differ by no more than $\sim 10\%$.  

An ideal way to compare the DNS with theory should be 
to find   $b(\rho)$  and $Pm_c$ in  the same DNS and to solve Eq.(\ref{eq-psi}) with this particular $b(\rho)$. 
Even when we neglect the term with $\sigma'$ (which is difficult to calculate numerically) in the potential, we would obtain a rather high accuracy of presumably better than 20\%. 
In this sense, the Kazantsev equation is a type of bridge  connecting  quasi-Lagrangian turbulence to the  parameters of the magnetic field generation. 

In absence of such a combined DNS, we took the data of \citet{biferale2011multi, donzis2010bottleneck} for the velocity statistics for one particular Reynolds number and compared the result with the data of \citet{schekochihin2007fluctuation}. The result is presented in Table~\ref{tab:2-140}, and the agreement is quite good.

One more difficulty for the comparison of the theory and the numerical/experimental  data is the  use of different normalizations by different authors.  This creates ambiguity and the additional need to coordinate  data. In order to avoid this, it would be better 
to  provide 
 the stability curve in terms of $Pm_c$ as a function of  $Re_{\lambda}$ , since, unlike   $Rm$ and $Re$, they are independent of the pumping properties of a particular flow and are universal in the sense that they can be expressed in terms of the basic  parameters $\varepsilon, v_{\rm rms}, \nu,\eta$.  
 Another way is to use a more universal definition of the integral scale. 
 We presented our results in terms of  $Re^{\rm Sch},\, Rm^{\rm Sch}$ that are defined by (\ref{Re-Schek}) and (\ref{defin-Rm}). 

The comparison of different models shows the importance of the transition region between the inertial and the integral ranges.  While the bottleneck transition between the viscous and inertial ranges is only important for relatively small Reynolds numbers,  this outer transition range was found to be essential for the magnetic field generation at all Reynolds numbers. 

For extremely high Reynolds numbers,  the piecewise power-law  model (Sharp) represents the qualitative properties of the generation well, but  lowers the generation threshold significantly;  the Smooth model appears to be much more accurate and agreed well with DNS.  For the 
velocity  scaling exponent  found by \citet{iyer2020scaling},   the obtained critical magnetic Reynolds number
fits the DNS results well.
 
The comparison of the Sharp and the Smooth models helped us 
to determine the region that is most important for the value of the generation threshold.   It is about twice smaller than the parameter $\Lambda$ of the Smooth model, which is comparable to  the pumping scale.
This scale belongs to the transition region from the inertial to the pumping range.  It can be marked by the condition  $\sigma (\rho)  \approx d \ln b/d \ln \rho =0  $, or by the claim that the effective  potential (\ref{U-pot})  is zero at the point. 

Based on the comparison of our data for different scaling exponents, we propose an explanation of the decrease in the critical magnetic Reynolds number as a function of $Re$ for $650\gtrsim Re_{\lambda} \gtrsim 300$. 
The earlier explanation referred to the effect of the bottleneck, but the downward tendency  of $Rm_c$ also continues for $Pm \lesssim 0.01$, where the effect of the bottleneck is  certainly negligible. 
We note that the critical magnetic Reynolds number depends essentially on the  scaling exponent of  the velocity structure function inside the inertial range.  As shown by \citet{iyer2020scaling}, the scaling exponent  decreases as a function of $Re$  from one-third at relatively small $Re$  to 0.39 as $Re \to \infty$.  This decrease is enough to  ensure the  decrease in $Rm_c$.  

We also studied the dependence of the growth rate exponent of the magnetic field correlator on the magnetic Reynolds number.   
The results are presented in Table~\ref{tab:5-gamma-compar}.
The theoretical prediction differs by two to four times from the numerical results, which in turn differ significantly from each other. On one hand, this is a good correspondence taking into account the technical uncertainties. On the other hand, the difference might be decreased when the non-Gaussianity of the velocity field is taken into account. As in the case of $Rm_c$,  the corresponding correction would change the theoretical result in the right direction. 

To conclude, we stress the importance of a precise comparison of the kinematic dynamo theory with DNS. It would quantitatively verify the key theoretical simplifications, such as a $\delta$-correlated in time effective velocity field. We obtained encouraging results, which confirm this assumption, 
by an attempt of such a comparison using the existing numerical and theoretical data on turbulence. 
We hope that future DNSs will be able to combine kinematic dynamo and Lagrangian velocity properties, which would  allow for an accurate comparison with the Kazantsev theory. 
The precise comparison  would also allow for a quantitative assessment of the effect of non-Gaussian effects, such as dynamo suppression and changes in the magnetic energy spectrum. A successful confirmation of the theory at moderately low magnetic Prandtl numbers would provide a critical basis for the extrapolation of  its predictions to extreme, astrophysically significant $Pm$ values, which are beyond  our
current computational capabilities. 

\begin{acknowledgements}
This work of AVK was supported by the Foundation for the Advancement of Theoretical Physics and Mathematics (BASIS).
\end{acknowledgements}

\bibliographystyle{aa} 
\bibliography{iucr} 

\begin{appendix}

\end{appendix}

\end{document}